# Rating the Online Restaurant Review Rating System using Yelp

Dhanasekar S, Balaji

SSN College, CEG Guindy

## ABSTRACT

The impact of ratings on a restaurant plays a major role in attracting future customers to that restaurant. The word of mouth has been systematically replaced with the online reviews. It gives a sense of satisfaction for people to know beforehand about the number of average stars the restaurant has acquired before stepping into a restaurant. However, these ratings are indirectly biased based on the location, amenities, and the perception of individual people. In this work, we analyze the ratings of restaurants available through the Yelp public data for the discrepancies in the rating system and attempt to provide an optimized global rating system. For a frequent visitor to a high-end restaurant with lavish amenities, even a slightest of reduction in the expected ambiance may prompt a 4-star rating, while a restaurant, which guarantees a minimum taste for its food, may get a 5-star rating. These discrepancies can often be attributed to three factors- the perspective of individual people, features of the restaurant and Location. The perspective of individual people is always subjective and what seems good for one person may be poor for another. In this work, we focus on the other two important factors – Reviews and the features. Handpicked restaurants with best amenities are used as a benchmark and the reviews of these restaurants are analyzed and scored for the appropriate star rating. Then the other restaurants other than that of the training set are optimized for their rating based on the 'true' rating learned through the training set. Experimental evaluation on hundreds of hotels with thousands of reviews give the optimized ratings updated on the original to indicate the true rating the restaurant deserves. This enables travelers and others who eat at different places to not get disappointed with the rating system through a false impression. E.g. A traveler who rates 5- stars for restaurants which offer many other factors other than good food like ambiance, courteous waitress and so on should not get a false impression on another restaurant which has a reputation of 5- stars solely based on the taste of food influenced by the location.

## 1 INTRODUCTION

In today's world, restaurants have become an integral part of everyday life. People often dine out more than ever. In such a setup, the ratings and reviews of a restaurant play a huge role in attracting future-to-be customers. One such online restaurant review system is Yelp [1]. It contains not only the ratings, reviews of any particular restaurant but also the list of amenities like parking, alcohol etc. A detailed description of the dataset and preprocessing is addressed in the next section. The central theme of this paper is identifying the disparity in the review ratings across different restaurants. As mentioned in the abstract, we identify these disparities using a combination of reviews and amenities. Using text mining, the review text is analyzed for sentiment scores, while the weighted amenities difference is the next parameter used to judge or compare any two restaurants. This ensures that review ratings of a restaurant are optimized based on the amenities and quality of reviews.

To summarize,

- First, we extracted data of over 50000 restaurants available in the Yelp dataset with features such as reviews, amenities.
- Second, we preprocessed and cleaned over 1M reviews of these restaurants posted by 200000 users.
- Third, the amenities were classified into four major categories based on the type of service it offers.
- Fourth, the category of features is weighted to signify the importance given to that amenity by users.
- Fifth, we sort restaurants based on the features and analyze their reviews to obtain sentiment score, which is used as a measure to compare and contrast other restaurants.

## 2 YELP DATASET

### 2.1 Dataset Extraction

The Yelp public dataset is available in three formats. We extracted Yelp SQL data comprising well over 6 Gigabytes. The extracted data is fed into MYSQL dump using Python. SQLAlchemy is the Python package used to analyze this SQL dump. The extraction and loading the data took well over 4 hours on a Mac book i7 with 12GB RAM.

## 2.2 Dataset Preprocessing

After extracting the data, the next important step is preprocessing. A separate table for users, business details, and reviews is created to assist the obtaining process. Each table is linked with a primary ID. The review table contains both the business and user IDs. The amenities of the restaurants were identified using flags provided by the Yelp. An attribute with 'Yes' or '1' indicates the presence of that feature in that particular restaurant as opposed to a 'No' or '0'. A separate amenity table was created which indicates the list of amenities offered by all the restaurants. Finally, a rank list of all the restaurants based on the number of amenities offered is created. The 55000 plus restaurants were sorted this way.

## 3 RELATED WORKS

There are a number of related works that have worked on restaurants review prediction and analysis. The first work explains hoe geographical neighborhood influence affects the rating prediction of any restaurants [2]. An excellent approach to combine rating system with the topic modeling from review texts is offered [3]. This work highly correlates with ours in that; we also employ the combination of review ratings and texts to analyze the disparities in ratings across restaurants. The next work [4] supports our argument of how the text reviews play a part in determining the review stars of the restaurant. Another work, which predicts the ratings of a restaurant based on the review texts alone, is also done. [5].

Another interesting work on how early ratings of a restaurant can often be misleading is detailed in [6]. It explains how the ratings of a restaurant usually tend to go down on an average and initial ratings are often high than the final one. Our work also can eliminate that bias by identifying true restaurants, which deserve these ratings. We take the restaurant reviews sentiment scores plus the features it has to offer, enabling to truly identify the rating of any restaurant automatically eliminating such bias.

Thus, we understand that [2] location makes an impact on how ratings are defined. Our work is partially inspired by [3,4,5] in that we use both review texts and ratings to derive a comprehensive system.

## 4 FEATURES OF A RESTAURANT

As described in the second section, the features of a restaurant are identified by mining the different attributes provided by the Yelp dataset for a restaurant. The features are divided into four different categories – Food, Parking, Amenities, and Qualities. Fig. 1 shows the various features of a restaurant classified under their types.

| Parking | Food | Amenities | Qualities |
|---|---|---|---|
| - BikeParking | - dessert | - HasTV | - RestaurantsGoodForGroups |
| - Garage | - latenight | - OutdoorSeating | - NoiseLevel |
| - Street | - Lunch | - BuisnessAccepts | - RestaurantsAttire |
| - Validated | - Dinner | - RestaurantsDelivery | - GoodForKids |
| - Lot | - Breakfast | - RestaurantsTakeOut | - classy |
| - Valet | - Brunch | - WiFi | - romantic |
| | - RestaurantsPriceRange2 | - RestaurantsTableService | - intimate |
| | - Alcohol | - RestaurantsCounterService | - hipster |
| | | - Caters | - touristy |
| | | - RestaurantsReservations | - trendy |
| | | | - upscale |
| | | | - casual |

**Figure 1: Features of a restaurant classified under different types.**

We then see the frequency of different features in Fig 2 for top 500 restaurants. We note how features diminish as we progress downwards after the top 500 in consecutive Figures 3 and 4.

| Features | Frequency |
|---|---|
| BusinessAcceptsCreditCards | 493 |
| RestaurantsGoodForGroups | 490 |
| RestaurantsTakeOut | 486 |
| RestaurantsPriceRange2 | 478 |
| RestaurantsTableService | 465 |
| NoiseLevel | 454 |
| Alcohol | 454 |
| HasTV | 438 |
| BikeParking | 436 |
| WiFi | 432 |

**Figure 2: Most frequent Features of top 500 restaurants**

| Features |
|---|
| RestaurantsPriceRange2 |
| BikeParking |
| HasTV |
| RestaurantsGoodForGroups |
| NoiseLevel |
| OutdoorSeating |
| RestaurantsTakeOut |
| WiFi |
| RestaurantsTableService |
| Alcohol |

| Features |
|---|
| RestaurantsPriceRange2 |
| BikeParking |
| OutdoorSeating |
| BusinessAcceptsCreditCards |
| RestaurantsTakeOut |
| WiFi |
| Caters |
| lot |

**Figure 3 a & b: Features of Restaurant no. 1000 and 10000 from top**

| Features |
|---|
| RestaurantsPriceRange2 |
| RestaurantsTakeOut |

**Figure 4: Features of Restaurant no. 50000 from top**

## 5   ESTIMATING THE SENTIMENT OF RESTAURANT REVIEWS

After analyzing all the features of all the restaurants, we sort all the restaurants based on the number of different features it offers. A list of top 500 restaurants which in the order of maximum features is built. That is the restaurant with most number of features stays at the top The next step is identifying the top words in the reviews of these restaurants. By top, the topics present in these reviews. We use the traditional TF-IDF approach for this and find all the topics in the reviews for each of the review ratings (1,2,3,4, and 5 stars) for all the 500 restaurants.

A complete list of topics for each restaurant, each review rating for 500 top restaurants is obtained. The sentiment scores of these topics founded using TF-IDF is obtained using [7]. Fig 5 shows the combined sentiment scores for the topic modeled review texts for top 500 featured restaurants (refer Section. 4).

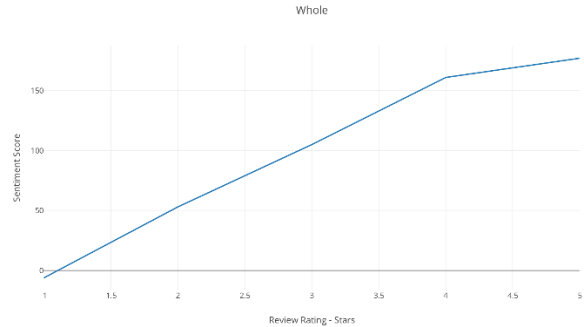

**Figure 5: Combined Sentiment scores for review texts of top 500 restaurants with most features.**

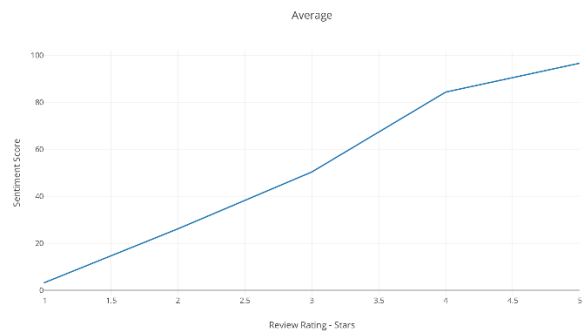

**Figure 6: Average Sentiment scores for review texts of top 500 restaurants with most features.**

Fig 5 gives us the information that on combining the review texts of all the individual ratings, we get a total sentiment score of *-6 for 1-star reviews text, around 50 for 2-star reviews text, 105 for 3 stars, 160 and 190 respectively for 4 and 5 stars*. This tells how dissatisfied customers were when they actually rate a 1-star review for a restaurant.

Fig. 6 tells about the average rating for the top 500 restaurants with most features. The range of scores is much more contracted with a *score of 3 for 1-star reviews text to all the way up to 97 for 5-star reviews*.

## 6   RESULTS- FINDING THE DISPARITY IN RATINGS

Now that we have the top restaurants in terms of features and their respective sentiment scores, we attempt to analyze the disparities between two restaurants with same overall ratings. We pick a random restaurant in our top- 500 lists and a one not in that list and start a detailed analysis.

We start by taking two restaurants with ids 'pHpU8lnnxMuPWRHOysuMIQ' and 'Vs7gc9EE3k9wARuUcN9piA'. The former one has an overall rating of 4.0 provided by Yelp and is classified as top- 500 by our classification. The second one has an overall rating of 4.5 but did not make it to the top- 500 list because of the difference gap in amenities.

We then find the common features between these two restaurants to be 'HasTV', 'RestaurantsPriceRange2', 'NoiseLevel', 'lot', 'RestaurantsTakeOut', 'BikeParking', 'BusinessAcceptsCreditCards', 'dinner', 'Caters', 'RestaurantsGoodForGroups', 'RestaurantsTableService', 'lunch', 'RestaurantsReservations', 'casual'. Then comes the interesting part. Even though the overall rating of the second restaurant is 4.5 greater than 4, the second restaurant has the following missing features - 'Alcohol', 'WiFi', 'OutdoorSeating', 'RestaurantsDelivery', 'brunch'. On the other hand, the first restaurant which we classified it as in top- 500 does not have any missing features compared to the restaurant with a rating star of 4.5. This tells us that the overall rating of the restaurant alone is not a clear measure to indicate the quality of that restaurant.

Further, to validate our claim, the sentiment scores of the review texts of these restaurants are higher for the former than the latter for few of the star ratings. The 3-star reviews sentiment score of restaurant 1 (90) is 1.69 times higher than the 3-star reviews sentiment score of restaurant 2, not classified as top-500. The 4-star reviews score is marginally higher for restaurant 1 than restaurant 2 like how it is for restaurant 2 in 5-star reviews.

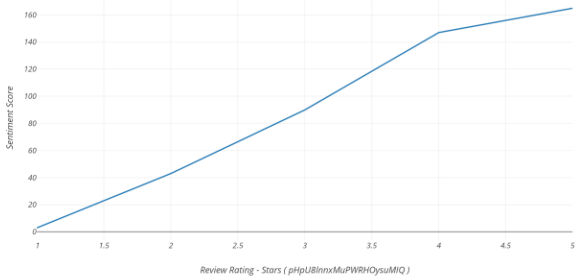

**Figure 7: Sentiment scores for review texts of restaurant 1.**

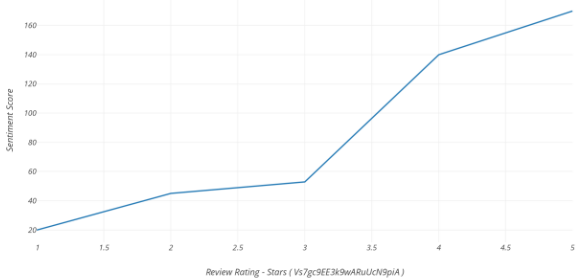

**Figure 8: Sentiment scores for review texts of restaurant 2.**

In fact, if we weigh each of the features by their type (E.g. 1.0 for 'Food', 0.8 for 'Parking', 0.7 for 'Amenity' and 0.6 for 'Quality') and measure the difference between these restaurants, we get a factor of 3.80 (weighted sum of deficient factors). One can assume the weights assigned as the importance given to each of the types of feature. The 'Food' category is assigned highest weight, as it is one of the primary features of a restaurant. The disparity can also be found using the difference in sentiment scores for each of the ratings. From the two graphs below, it is evident that the restaurant 1 with 4.0 is favored in sentiment scores also. *The net sentiment score is -17-2+37+7-5, which is 20*.

In the end, our analysis, both through sentiment score and weighted features difference favors the restaurant with a rating 4.0 over another restaurant with a rating 4.5. A detailed result consisting of tables and graphs is extensively provided in [8].

## 7 CONCLUSION

As stated in the abstract, we were able to identify the disparities in standards of reviews across any two given restaurants using online restaurant review system Yelp. The sentiment score-based analysis helped identify the difference using reviews text while the weighted feature-based analysis sorted out the restaurants based on the various features offered. In the future, we would like to bring in the person based analysis also into the picture as reviews across multiple restaurants may be influenced by a certain group of people, which may make the analysis much more comprehensive.